# Green Cellular Network Deployment

# To Reduce RF Pollution

**Sumit Katiyar**
Research Scholar
Singhania University
Jhunjhunu, Rajasthan, India
sumitkatiyar@gmail.com

**Prof. R. K. Jain**
Research Scholar
Singhania University
Jhunjhunu, Rajasthan, India
rkjain_iti@rediffmail.com

**Prof. N. K. Agrawal**
Sr Member IEEE
Inderprastha Engg. College
Sahibabad, Ghaziabad, India
agrawalnawal@gmail.com

## ABSTRACT
As the mobile telecommunication systems are growing tremendously all over the world, the numbers of handheld and base stations are also rapidly growing and it became very popular to see these base stations distributed everywhere in the neighborhood and on roof tops which has caused a considerable amount of panic to the public in Palestine concerning wither the radiated electromagnetic fields from these base stations may cause any health effect or hazard. Recently UP High Court in India ordered for removal of BTS towers from residential area, it has created panic among cellular communication network designers too. Green cellular networks could be a solution for the above problem. This paper deals with green cellular networks with the help of multi-layer overlaid hierarchical structure (macro / micro / pico / femto cells). Macrocell for area coverage, micro for pedestrian and a slow moving traffic while pico for indoor use and femto for individual high capacity users. This could be the answer of the problem of energy conservation and enhancement of spectral density also.

## Keywords
Smart / Adaptive Antenna, Picocell, Femtocell, Remote RF, Hierarchical Structure

## 1. INTRODUCTION
Across the developed and developing world, wireless communications has proven a necessity. There seems no end in sight to the proliferation of mobile communications as over 120,000 new base stations are deployed yearly, and there is growth across every demographic from teenagers to businessmen to the poorest Indian village. The developing world has turned to wireless communications as a leap frog technology past wired communications which spurs its growth even more. Simultaneously, the industrialized world has developed an insatiable demand for broadband data (including internet and multimedia services) delivered through their cellular handset. This meteoric rise in users and data demand alone does not create a crisis; however when one evaluates the communications ecosystem from a carbon footprint and energy cost perspective, the results are startling. A medium sized cellular network uses as much energy as 170,000 homes. While the cost of powering the needed base stations accounts for a staggering 50% of a service provider's overall expenses. This impact is magnified by the requirement for expensive 'dirty' diesel fuel for many locations in developing regions. The deteriorating economic landscape combined with the emerging emphasis on stewardship to the environment has made the 'greening' of communications an imperative.

R.F. pollution is harmful to mankind and can cause neurological, cardiac, respiratory, ophthalmological, dermatological and other conditions ranging in severity from headaches, fatigue and add to pneumonia, psychosis and strokes.

This paper provides the probable solution for keeping RF pollution level with in harmless limits.

Rising energy costs and the recent international focus on climate change issues has resulted in a high interest in improving the energy efficiency in the telecommunications industry. In this work, the effects of a joint deployment of macrocells / microcells for area coverage and publicly accessible residential picocells / femtocells on the total energy consumption of the network are investigated. The decreasing energy efficiency of today's macrocellular technologies with increasing user demand for high data rates is also discussed. It is shown that a joint deployment of macro, micro- and publicly accessible residential picocells / femtocells can reduce the total energy consumption by up to 60% in an urban area with today's technology [1]. The macrocell are utilized for high speed traffic, microcell for low speed traffic and pedestrian use. However picocells / femtocells are for indoor traffic and publicly accessible areas (hotspots). Furthermore the impact of future technologies for both macro- and femtocells on the energy efficiency is investigated [2]. It is shown that the benefits of a joint macro- and picocell deployment will increase further as both technologies mature, and will result in a significant reduction of the network energy consumption as the user demand for high data rates increases.

Energy efficiency in cellular networks is a growing concern for cellular operators to not only maintain profitability, but also to reduce the overall environment effects. This emerging trend of achieving energy efficiency in cellular networks is motivating the standardization authorities and network operators to continuously explore future technologies in order to bring improvements in the entire network infrastructure [2]. In this work, we present a brief survey of methods to improve the power efficiency of cellular networks, explore some research issues and challenges and suggest some techniques to enable an energy efficient or "green" cellular network. Since base stations consume a maximum portion of the total energy used in a cellular system, we will first provide a comprehensive survey on techniques to obtain energy savings in base stations. Next, we discuss how heterogeneous network deployment based on micro, pico and femtocells can be used to achieve this goal. Since cognitive radio and cooperative relaying are undisputed future technologies in this regard, we propose a research vision to make these technologies more energy efficient for future. Lastly, we explore some broader perspectives in realizing a "green" cellular network technology.

In this paper we have restricted ourselves with the up gradation of existing 2G / 2.5G networks for achieving our target of green cellular networks.

## 2. EFFECTS OF JOINT MACROCELL / MICROCELL AND RESIDENTIAL PICOCELL / FEMTOCELL FOR DEPLOYMENT OF GREEN CELLULAR NETWORKS

The requirements of green cellular network can be fulfilled with the known cellular architectures (macro, micro, pico cells) led to the conception of the idea of a hierarchical cell structure [3],[4]. The key issue for this type of cell architecture is to apply multiple cell layers to each service area, with the size of each layered cell tailored to match the required traffic demand and environmental constraints (Fig. 1). In essence, microcells will provide the basic radio coverage but they will be overlaid with Umbrella cells to maintain the ubiquitous and continuous coverage required. Especially for the DS-CDMA system, this mixed cell technique gives answers to situations where possible performance degradation may occur, e.g. fast moving users requiring handover, or black spots in coverage.

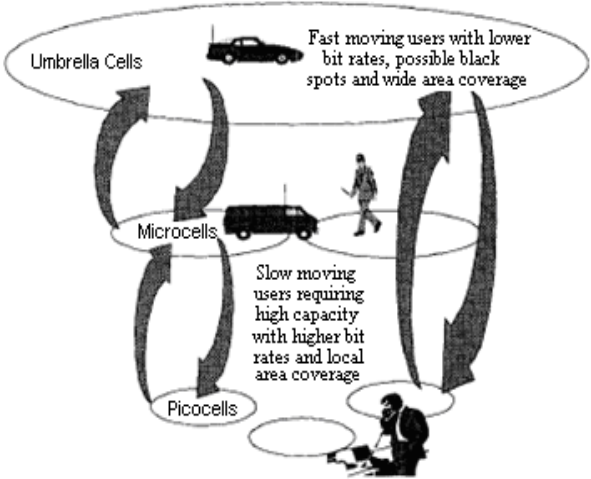

**Fig 1: Hierarchical Cell Structure Concept [9]**

Advanced antenna techniques, such as adaptive antennas, are an area which seems to gather momentum recently [5], [6], [7], [8] as another possible way to increase the efficiency of a given system. Adaptive antennas, based on the spatial filtering at the base station, separate the spectrally and temporally overlapping signals from multiple mobile units. This can be exploited in many ways such as:

- Support a mixed architecture.
- Combat the near-far effect.
- Support higher data rates.
- Combine all the available received energy, (multipath).

Residential picocells are deployed in conjunction with a wide area cellular network for area coverage in an urban environment as illustrated in fig. 2. It is assumed that the residential picocells have substantial auto-configuration capabilities to support simple plug and play deployment by customers.

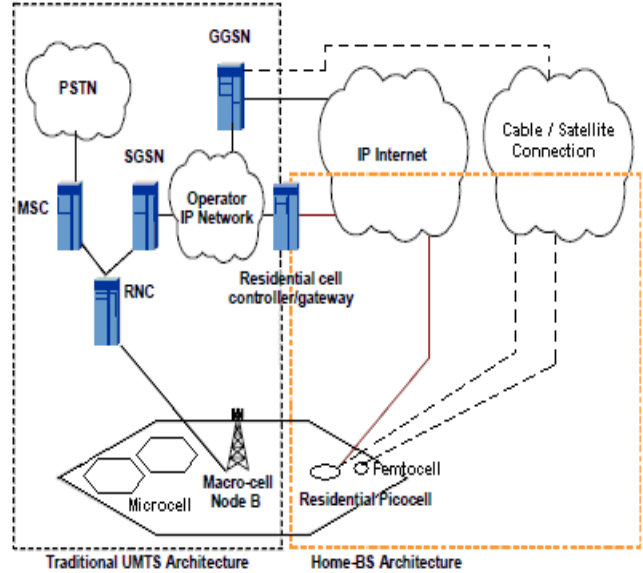

**Fig 2: Overview of macrocellular underlay network with residential picocell overlay deployment**

## 3. AN ADAPTIVE BASE STATION ANTENNA FOR THE UMBRELLA CELL OF A MIXED CELL STRUCTURE

The potential advantages offered by employing an adaptive antenna at an Umbrella BS site with a DSCDMA system, can be summarized as follows:

- Mitigation of the near-far effect.
- Capacity enhancement
- More efficient handover.
- "In fill" coverage for the dead-spots.
- Ability to support high data rates.

## 4. ADAPTIVE SPATIAL DIVISION MULTIPLE ACCESS

A-SDMA is based on the exploitation of the spatial dimension which has so far not been used for parallelism. Using adaptive array antennas at the base station sites, multiple independent beams can be formed with which several users can be served simultaneously on the same radio channel. This approach is an extension of the system where adaptive arrays are used for interference reduction, without exploiting the potential of spatial parallelism. This is illustrated in fig 3. Benefits of A-SDMA are proven and have been demonstrated in [11]. The benefits of adaptive antenna are also proven and can be easily implemented [10]. *In essence, the scheme can adapt the frequency allocations to where the most users are located.*

Because SDMA employs spatially selective transmission, an SDMA base station radiates much less total power than a conventional base station. One result is a reduction in network-wide RF pollution. Another is a reduction in power amplifier size. First, the power is divided among the elements, and then the power to each element is reduced because the energy is being delivered directionally. With a ten-element array, the amplifiers at each element need only transmit one-hundredth the power that would be transmitted from the corresponding single antenna system [12].

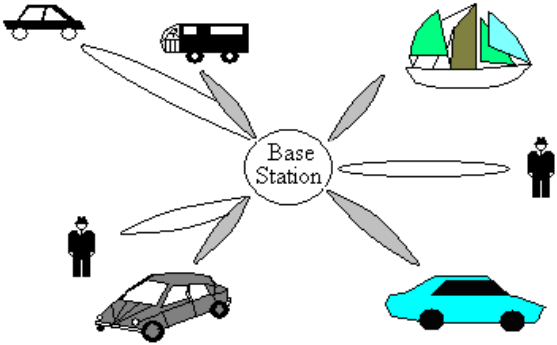

**Fig 3: Typical scenario for A-SDMA application**

## 5. HIERARCHICAL STRUCTURE

Cellular networks are becoming increasingly heterogeneous due to the co-deployment of many disparate infrastructure elements, including micro, pico and femtocells, and distributed antennas. A flexible, accurate and tractable model for a general downlink HCN consisting of K tiers of randomly located BSs, where each tier may differ in terms of average transmit power, supported data rate, and BS density. Assuming 1) a mobile connects to the strongest BS, 2) the target Signal to- Interference-Ratio (SIR) is greater than 0 dB, and 3) received power is subject to Rayleigh fading and path loss. Expressions for the average rate achievable by different mobile users are derived. This model reinforces the usefulness of random spatial models in the analysis and research of cellular networks. This is a baseline tractable HCN model with possible future extensions being the inclusion of antenna sectoring, frequency reuse, power control and interference avoidance/ cancellation [13]. To overcome handoff problem in hierarchical cell structure, efficient use of radio resources is very important. All resources have to be optimally utilized. However, in order to adapt to changes of traffic, it is necessary to consider adaptive radio resource management.

The ability of hierarchical cellular structure (Fig 3) with inter-layer reuse to increase the capacity of a mobile communication radio network by applying Total Frequency Hopping (T-FH) and Adaptive Frequency Allocation (AFA) as a strategy to reuse the macro- and micro cell resources without frequency planning in indoor picocells / femtocells have been discussed [14].

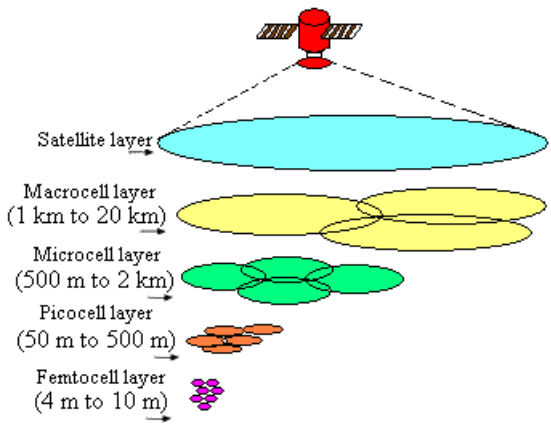

**Fig 4: Hierarchical Cellular Structure**

### 5.1 Macro Cell
A conventional base station with 20W power and range is about 1 km to 20 km. Macro cell in hierarchical structure takes care of roaming mobiles.

### 5.2 Micro Cell
A conventional base station with 1W to 5W power and range is about 500 m to 2 km. Micro cells and pico cells takes care of slow traffic (pedestrian and in-building subscribers). Micro cells can be classified as the following:

#### 5.2.1 Hot Spots
These are service areas with a higher tele-traffic density or areas that are poorly covered. A hot spot is typically isolated and embedded in a cluster of larger cells.

#### 5.2.2 Downtown Clustered Micro cells
These occur in a dense, contiguous area that serves pedestrians and mobiles. They are typically found in an "urban maze of" street canyons," with antennas located far below building height.

#### 5.2.3 In-Building. 3-D Cells
These serve office buildings and pedestrians (fig 4). This environment is highly clutter dominated, with an extremely high density and relatively slow user motion and a strong concern for the power consumption of the portable units.

### 5.3 Pico Cell
The picocells are small versions of base stations, ranging in size from a laptop computer to a suitcase. Besides plugging coverage holes, picocells are frequently used to add voice and data capacity, something that repeater and distributed antenna cannot do.

Adding capacity in dense area, splitting cells are expensive, time consuming and occasionally impossible in dense urban environment where room for a full size base station often is expensive or unviable. Compact size picocells makes them a good fit for the places needing enhanced capacity, they can get.

Picocells are designed to serve very small area such as part of a building, a street corner, malls, railway station etc. These are used to extend coverage to indoor area where outdoor signals do not reach well or to add network capacity in areas with very dense uses.

### 5.4 Femto Cell
A femtocell is a smaller base station, typically designed for use in home or small business**.** In telecommunications, a femtocell is a small cellular base station, typically designed for use in a home or small business. It connects to the service provider's network via broadband (such as DSL or cable).

#### 5.4.1 Advantage to the Femtocell User

##### 5.4.1.1 Better coverage and capacity
Due to their short transmit-receive distance, femtocells can greatly lower transmit power, prolong handset battery life, and achieve a higher signal-to interference-plus-noise ratio (SINR). These translate into improved reception—the so-called five-bar coverage—and higher capacity. Because of the reduced interference, more users can be packed into a given area in the same region of spectrum, thus increasing the area spectral efficiency [15], or equivalently, the total number of active users per Hz per unit area.

*5.4.1.2 Improved macrocell reliability*

If the traffic originating indoors can be absorbed into the femtocell networks over the IP backbone, the macrocell BS can redirect its resources towards providing better reception for mobile users.

*5.4.1.3 Cost benefits*

Femtocell deployments will reduce the operating and capital expenditure costs for operators. A typical urban macrocell costs upwards of $1K per month in site lease, and additional costs for electricity and backhaul. The macrocell network will be stressed by the operating expenses, especially when the subscriber growth does not match the increased demand for data traffic. The deployment of femtocells will reduce the need for adding macro-BS towers. A recent study [16] shows that the operating expenses scale from $60K per year per macrocell to just $200 per year per femtocell.

*5.4.1.4 Reduced subscriber turnover*

Poor in-building coverage causes customer dissatisfaction, encouraging them to either switch operators or maintain a separate wired line whenever indoors. The enhanced home coverage provided by femtocells will reduce motivation for home users to switch carriers.

The goal of this paper provided an overview for these end-user deployed infrastructure enhancements, and describe in more detail how the above improvements come about. We also describe the business and technical challenges that femtocells present, and provide some ideas about how to address them.

To summarize, the capacity benefits of femtocells are attributed to:

- Reduced distance between the femtocell and the user, which leads to a higher received signal strength.
- Lowered transmit power, and mitigation of interference from neighboring macrocell and femtocell users due to outdoor propagation and penetration losses.
- As femtocells serve only around 1-4 users, they can devote a larger portion of their resources (transmit power & bandwidth) to each subscriber. A macrocell, on the other hand, has a larger coverage area (500m-1 km radius), and a larger number of users; providing Quality of Service (QoS) for data users is more difficult.

# 6. REMOTE RF TECHNOLOGY

Global warming has become an increasingly important item on the global political agenda. In this regard, information in communication technologies have been identified to be a major future contributor to overall green house gas emissions, having a share of more than 2% already in 2007 with a strong trend to increase [17] [18]. In the strive for lessening of the environmental impact on the information and communication industry, energy consumption and green house gas emissions of communication networks has recently received increased attention.

The technologies and applications associated with the extension of connection between a wireless radio base station and its antennas can be a probable solution to overcome this problem. The extension of this connection allows wireless operators to provide wireless service to new and remote areas, using radio equipment from an existing location. Optical Wireless Technology is proposed as a viable and cost effective technology for this application in urban environments. The cost benefits of using today's technologies to extend the link between radio base stations and their antennas is also explained.

Distributed cell site architecture will be explored as a possible RF network design concept. The potential benefits of this concept will be described in detail.

The benefits of cell site deployment using the distribution of centralized radio resources are listed below:

- Significant cost savings achieved by aggregating backhaul traffic for the hub and remote cells – hub provides shared backhaul facilities.
- Costs associated with leasing building space are significantly reduced, since indoor building space will not be required for most remote locations.
- Reduced deployment and operational costs associated with providing cell site building facilities such as power, electricity, air conditioning, fire protection and security.
- Remote sites can be deployed much more quickly than traditional cell sites.
- Lower overall infrastructure installation costs resulting from base station co-location at hub cells.
- Lower maintenance and operating expenses due to base station collocation at hub cells.
- Enhanced deployment flexibility with remote cells.
- Lower overall deployment costs for new networks.
- Re-use of current cell site facilities for existing networks.
- Facilitates cost effective in-building coverage and capacity solutions.

This method of cell site deployment is even more powerful when it is combined with the sharing of facilities and equipment between wireless operators. Some or all of the remote RF technologies may be used to facilitate the distribution of centralized radio resources. It is further suggested that Optical Wireless Technology may be ideal for this application, for dense urban areas in particular. The equipment can be mounted on rooftops with the RF antennas, and there are no spectrum licensing requirements. It provides the capacity required for dense urban environments, and it is also easy to install and maintain. Optical Wireless Technology could help operators deploy cells much more quickly, which would improve their quality of service and customer satisfaction.

In many urban cell site locations, more than one operator has equipment at the site. Some sites may have multiple wireless service providers operating from the same location. In some of these locations, operators share the antennas, cable and backhaul transport facilities. As the trend for sharing facilities and wireless equipment continues, optical technology will provide the capacity required to combine backhaul and other facilities for multiple operators. Operators can use all of these techniques and a variety of remote RF technologies to reduce the costs of network deployment. Overall, there is a compelling business case for the use of remote RF technology to build and expand wireless networks.

Remote RF solutions have evolved as wireless network infrastructure has evolved. A variety of remote RF solutions exist today to help operators provide improved capacity and coverage. These solutions can be implemented individually or in tandem to achieve the desired results. Optical Wireless Technology is viable option for remote RF applications in urban environments.

Wireless operators can generate tremendous cost savings by utilizing remote RF technology to distribute centralized radio resources. The operator's savings are increased significantly as an increasing number of remote cells are added in lieu of conventional cells. The implementation of remote cells provides reductions in both the capital and operating expenses associated with new cell site deployment. Wireless operators should consider the implementation of remote cells as a part of their overall network deployment, expansion of their existing network, or for the implementation of new wireless access technologies.

## 7. PROPOSED GREEN CELLULAR NETWORK

In the prevailing scenario the existing 2G/2.5G network will continue to dominate in India. We have to address the problem of spectral density for catering the growing demand as well as RF pollution and energy conservation. In this paper a simple solution has been suggested to meet the above requirements for up gradation of existing 2G/2.5G network in India.

The proposed evolution path for green network is given in fig 5.

Conventional / sectored antennas are replaced by intelligent / adaptive antenna along with A-SDMA approach. Picocells and femtocells are replaced instead of conventional distributed networks. The use of adaptive antenna will not only reduce the power consumption but also reduce RF pollution drastically in the existing system. In addition to it the growing demand will also be met. Indoor users / high capacity in-building and home users will be covered through deployment of low power pico cell / femtocell (fig. 6).

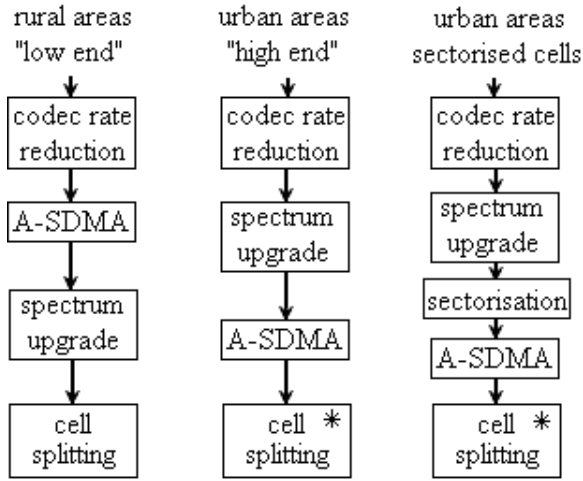

**Fig 5: Outage probability with sectorized antenna**

Small power pico and femto cells are extensively used in proposed network instead of distributed networks for in-building and high capacity indoor users will be provided with better quality of service in comparison to other indoor existing available technologies.

High capacity indoor users have been covered with low power femto cell while hotspots and indoor users will be covered with low power picocell instead of distributed networks to provide much better QoS in comparison to existing wireless technologies for indoor subscribers.

Use of adaptive antenna, low power pico cell and femto cell in existing 2G / 2.5G networks not only meet the increasing demand but also reduce RF pollution and power consumption drastically with out any doubt.

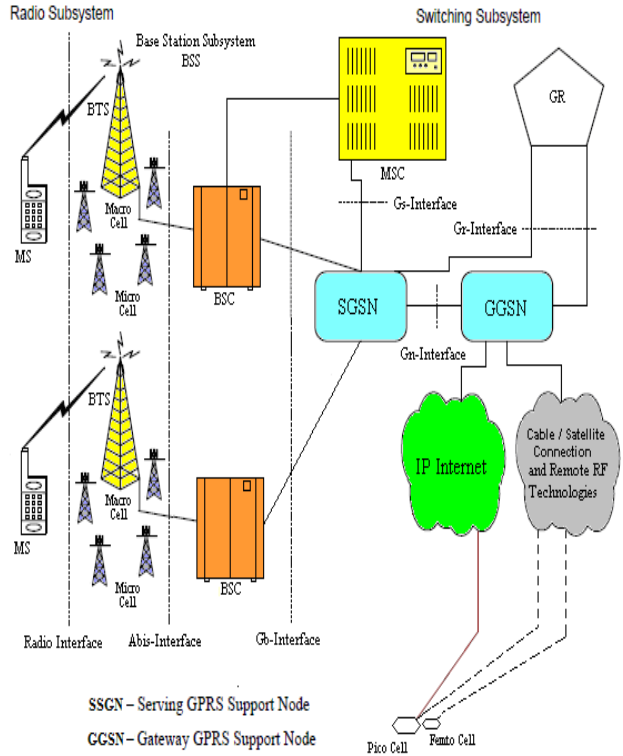

**Fig 6: Proposed Green Cellular Network**

## 8. CONCLUSION

It has been proved beyond doubt that incorporation of intelligent / adaptive antenna along with A-SDMA approach alone can enhance spectral density, reduce power consumption and RF pollution drastically in cellular communication network [1], [19], [20]. In addition to it inclusion of remote RF technology will further enhance capacity of the network. As existing systems optical network, wireless network etc. will be used in RF remote technology. The power consumption as well as RF pollution will reduce further. Moreover in this paper, extensive use of picocell and femtocell will further reduce power consumption and RF pollution drastically, as some conventional BTS will be replaced with low power picocell & femtocells which will further reduce power consumption and RF pollution.